\documentclass[a4paper]{jpconf}
\usepackage{graphicx}
\usepackage{iopams}
\begin{document}
\title{Ab Initio Nuclear Structure Calculations of Light Nuclei}

\author{Pieter Maris}

\address{Department of Physics and Astronomy, 
Iowa State University, Ames, IA 50011, USA}

\ead{pmaris@iastate.edu}

\begin{abstract}
We perform no-core configuration interaction calculations for nuclei
in the $p$-shell.  We show that for typical light nuclei, a truncation
on the total number of quanta in the many-body system converges much
more rapidly than a full configuration interaction (FCI) truncation,
which is a truncation on the single-particle basis space.  We present
new results for the ground state energies of the Be isotopes with the
nonlocal two-body potential JISP16, and discuss emerging phenomena
such as clustering and rotational band structures in $^9$Be.  We also
show that the anomalously suppressed beta decay of $^{14}$C to the
ground state of $^{14}$N can be reproduced using two- and
three-nucleon forces from chiral effective field theory.  In
particular the structure of the ground state of $^{14}$N is sensitive
to the three-nucleon force.
\end{abstract}

\section{Ab initio nuclear physics}

Solving for nuclear properties with the best available nucleon-nucleon
(NN) potentials~\cite{Wiringa:1994wb,Entem:2003ft}, supplemented by
three-nucleon forces (3NF) as
needed~\cite{Pieper:2001ap,Epelbaum:2005pn,Navratil:2007we}, using a
quantum many-particle framework that respects all the known symmetries
of the potentials is referred to as an "ab initio" problem and is
recognized to be computationally hard.  Through the UNEDF SciDAC
collaboration~\cite{bertsch2007,furnstahl2011,nam2011} of nuclear
theorists, applied mathematicians, and computer scientists there has
been a rapid development of ab initio methods for solving finite
nuclei, which has opened a range of nuclear phenomena that can now be
evaluated to high precision using realistic inter-nucleon
interactions.

A commonly used approach in nuclear physics is the Configuration
Interaction (CI) method for solving the many-body nuclear Hamiltonian
in a (sufficiently large) single-particle basis space.  In this approach,
the many-body Schrodinger equation becomes a large sparse matrix problem.
The eigenvalues of this matrix are the binding energies, and the
corresponding eigenvectors the nuclear wave functions, which can be
employed to evaluate experimental quantities.  In order to reach
numerical convergence for fundamental problems of interest, the matrix
dimension often exceeds two billion; obtaining the lowest eigenvalues
and eigenvectors of such large sparse matrices poses significant
challenges even on today's leadership class facilities.

In this work, we discuss convergence rates of CI calculations,
contrasting Full Configuration Interaction (FCI), which is a
truncation on the single-particle basis space only, with the
$N_{\max}$ truncation, which is a truncation on the many-body basis
space.  The latter is commonly used in nuclear structure calculations,
not only because of its superior convergence rate for light nuclei,
but also because it leads to an exact seperation of the center-of-mass
(cm) motion and the translationally-invariant wavefunction.

We use a recently introduced technique for unfolding the cm motion
from the one-body density matrices~\cite{cockrell2012}, which allows
us to obtain translationally-invariant nucleon densities.  With this
technique we can reveal interesting phenomena such as clustering
emerging from our ab initio caculations.  In particular we show that
the ground state of $^9$Be consists of two clusters of protons and
neutrons separated by a few fm, consistent with $\alpha$-cluster
models, with the 'extra' neutron forming a ring or donut shape.  We
also discuss emerging rotational band structure in $^9$Be, and present
new results for the ground state energies of the beryllium isotopes.

Finally, we discuss recent results for the anomalously long lifetime
of $^{14}C$, which can be reproduced by the inclusion of three-nucleon
forces.  We analyse in detail what the influence of the 3NF is, not
only on the transition between the ground states of $^{14}$C and
$^{14}$N, but also on the transition between the first excited $2^+$
state of $^{14}$C and the ground state of $^{14}$N.  We identify that
it is the structure of the ground state of $^{14}$N, rather than that
of $^{14}$C, that responds sensitively to the 3NF, and therefore leads
to the anomalously small transition matrix element.

\section{No-Core Configuration Interaction calculations of light nuclei}

In a no-core CI approach, the wavefunction $\Psi$ of a nucleus
consisting of $A$ nucleons (protons and neutrons) is expanded in an
$A$-body basis of Slater Determinants of single-particle states; that
is
\begin{eqnarray}
  \Psi(\vec{r_1}, \ldots, \vec{r_A}) &=& 
    \sum c_k \Phi_k(\vec{r_1}, \ldots, \vec{r_A})  \,,
\label{Eq:expansion}
\end{eqnarray}
where $\Phi_k$ are $A$-body Slater Determinants.  Conventionally, one
uses a harmonic oscillator (HO) basis for the single-particle states,
but it is straightforward to extend this approach to a more general
single-particle basis~\cite{caprio2012CS}.

We use the so-called $M$-scheme for our basis.  The single-particle
states are labelled by the quantum numbers $n$, $l$, $j$, and $m_j$,
where $n$ and $l$ are the radial and orbital HO quantum numbers, with
$N=2n+l$ the number of HO quanta; $j$ is the total single-particle
spin, and $m_j$ its projection along the $z$-axis.  The many-body
basis states have well-defined total spin-projection, which is simply
the sum of $m_j$ of the single-particle states, $M_j = \sum m_j$,
hence the name $M$-scheme.  However, in general the many-body basis
states do not have a well-defined total $J$.  Some of the advantages
of this scheme is that it is very simple to implement, and that in two
runs (for positive and negative parity), we get the complete low-lying
spectrum, including the ground state, even if we do not know what the
spin of the ground state is.

The interaction $V$, typically a two-body (NN), sometimes supplemented
by a three-body (3NF) potential, is also expressed in this many-body
basis, as is the kinetic energy $T$.  Diagonalizing the resulting
matrix $H = T + V$ gives us the eigenstates and eigenenergies.  In
general, we are only interested in the low-lying spectrum, so we do
not need a complete diagonalization of this matrix, which can be
rather large, exceeding 12 billion basis states in a recent
application.

\subsection{Truncation and Convergence}

In principle, by expanding the nuclear wavefunction in a complete
basis, we would obtain exact results for a given input interaction
$V$.  However, such a basis is infinite-dimensional, and practical
calculations can only be done in a finite-dimensional truncation of a
complete basis.  In order to recover exact results, independent of the
basis space truncation, we typically perform a series of calculations
in increasingly large basis spaces.  Based on the results of such a
series of calculations, we either find that we have achieved numerical
convergence (within a certain numerical uncertainty), or extrapolate
to the complete basis (with a certain extrapolation uncertainty).  Of
course, for some observables it is much harder to achieve convergence
than for others, as we shall see.  In addition, one could employ
renormalization techniques to accelerate convergence in relatively
small basis spaces.

Full Configuration Interaction or FCI employs a widely-know truncation
in which all many-body basis states are retained that can be
constructed from a finite set of single-particle states.  This method
is commonly used in atomic and molecular physics.

In ab initio nuclear physics however, we generally use a
truncation on the many-body basis space, rather than on the
single-particle basis.  To be specific, we construct a
finite-dimensional basis by a truncation on the total number of quanta
of the system.  That is, the many-body basis for a nucleus consisting
of $A$ nucleons is limited to states that satisfy the condition
\begin{eqnarray}
  \sum_{k = 1}^A N_k & \le & N_0 + N_{\max} \,,
\end{eqnarray}
where $N_k$ is the number of quanta of each of the single-particle
states in the many-body basis state; $N_0$ is the minimal number of
quanta for that nucleus; and $N_{\max}$ is the truncation parameter.
For HO single-particle states, or more general, for any basis in which
the single-particle states have radial and orbital quantum numbers $n$
and $l$, we have $N_k = 2 n_{k} + l_{k}$.

There are several advantages to this $N_{\max}$ truncation.  One is
that, in combination with a HO basis for the single-particle basis, it
leads to an exact factorization of the cm motion and the relative
motion.  In nuclear physics, we deal with a selfbound system, and we
have to either use relative coordinates for our calculations, or (as
we do here) single-particle coordinates.  Note that the wavefunctions
in Eq.~(\ref{Eq:expansion}) are functions of $A$ (independent)
single-particle coordinates, not $(A-1)$ relative coordinates.  This
makes the anti-symmetrizations and bookkeeping in the numerical codes
almost trivial, but it does mean that we have to somehow seperate out
the cm wavefunction from the translationally-invariant wavefunction.
A HO single-particle basis, in combination with $N_{\max}$ truncation,
allows for an exact factorization of the cm wavefunction
\begin{eqnarray}
  \Psi(\vec{r_i}) &=& \Psi^\omega_{\rm cm}(\vec{R}) \otimes \Psi_{\rm ti} \,,
\label{Eq:exactfact}
\end{eqnarray}
where $\vec{R}=(\frac{1}{A})\sum_{i=1}^A\vec{r}_i$ and $\Psi_{\rm ti}$
does not depend on the cm motion.  In order to remove the states with
cm excitations from the low-lying spectrum, we use the Lawson
method~\cite{Gloeckner1974313} whereby we add a Lagrange multiplier
term, $\lambda(H^\omega_{\rm cm} - \frac{3}{2}\hbar\omega)$, to the
many-body Hamiltonian.  The actual Hamiltonian that we are using thus
takes the form
\begin{eqnarray}
  H^{\lambda, \omega} &=&
  T_{\hbox{\scriptsize rel}} + \lambda_{\rm cm} 
 \bigg( H^{\rm HO}_{\rm cm} - \frac{3}{2} \hbar \omega \bigg)
  + \sum_{i<j} V_{ij} + \sum_{i<j<k} V_{ijk} + \ldots
\end{eqnarray}
Note that although this Hamiltonian depends explicitly on the HO
parameter $\hbar\omega$, as well as on $\lambda$, the low-lying
spectrum (i.e. states with $E_{\hbox{\scriptsize excitation}} \lesssim
\lambda \hbar\omega$) only depends implicitly on $\hbar\omega$ through
the basis space truncation, and becomes independent of $\hbar\omega$
in the infinite-basis limit.

\begin{figure}[t]
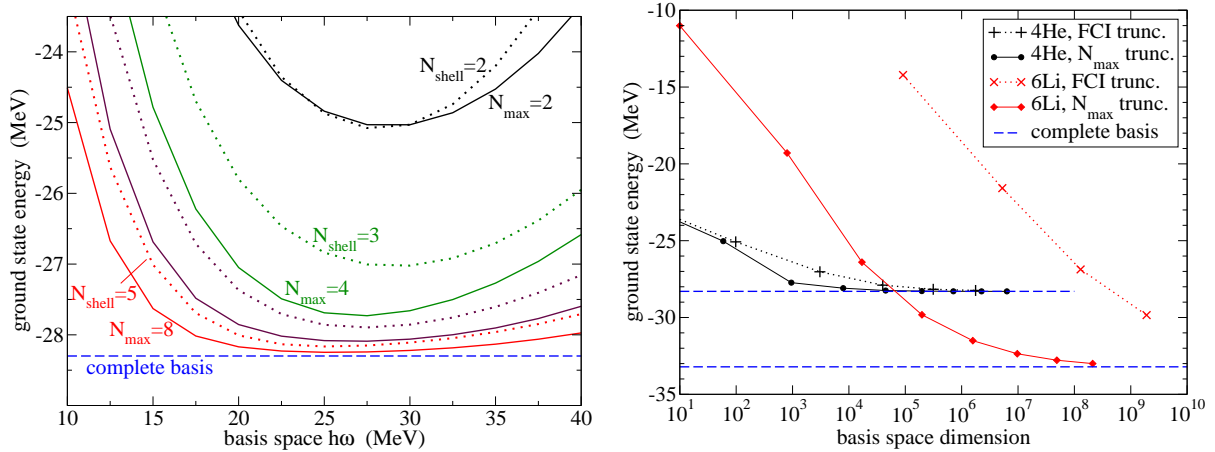

\includegraphics[width=0.48\textwidth]{results_4He.eps}\quad
\includegraphics[width=0.48\textwidth]{NCSMvsFCI.eps}
\caption{\label{Fig:FCIvsNmax} 
  Comparison of the convergence rates with JISP16 
  using FCI (dotted) and $N_{\max}$ truncations (solid).  
  Left: the ground state energy of $^4$He 
  as function of $\hbar\omega$; 
  Right: the ground state energies of $^4$He and $^6$Li 
  as function of the basis space dimension.
  The dashed line represents the exact result in a complete basis.}
\end{figure}
In the infinite-basis limit both FCI and the $N_{\max}$ truncations
should converge to the same exact result.  It turns out that the
numerical convergence is significantly faster with an $N_{\max}$
truncation than with an FCI truncation~\cite{Abe:2011pk,Abe:2012wp} in
our nuclear physics applications.  In Fig.~\ref{Fig:FCIvsNmax} we
compare the convergence rate for two light nuclei, $^4$He and $^6$Li,
with an FCI truncation and with an $N_{\max}$ truncation.  These
calculations were performed with a realistic two-body interaction
derived from inverse scattering theory, JISP16~\cite{Shirokov:2005bk},
which has been demonstrated to have good convergence rates for the
ground state energies of nuclei with $A \leq16$~\cite{Maris:2008ax};
and it gives a good description of light nuclei without explicit
three-body forces.

The left panel of Fig.~\ref{Fig:FCIvsNmax} shows that for $^4$He both
FCI and the $N_{\max}$ truncation have a similar convergence rate, at
least in terms of the HO parameter $\hbar\omega$ and the truncation
parameter $N_{\hbox{\scriptsize shell}}$ and $N_{\max}$ respectively.
As the truncation parameter increases, the dependence on $\hbar\omega$
decreases, and the succesive $N_{\hbox{\scriptsize shell}}$ curves
(dotted) and $N_{\max}$ curves (solid) converge to the same ground
state energy of $-28.299$~MeV.

However, as function of the basis space dimension, the $N_{\max}$
truncation converges much more rapidly than the FCI calculations, as
is illustrated in the right panel of Fig.~\ref{Fig:FCIvsNmax}.  E.g.,
an $N_{\max}=8$ calculation with a basis dimension of less than 50,000
for $^4$He leads to a ground state energy that is well within 100 keV
of the exact result, whereas for an FCI calculation
$N_{\hbox{\scriptsize shell}}=6$ with a basis dimension of almost 2
million is needed in order to achieve a similar level of convergence.
For $^6$Li, an FCI calculation with a basis dimension of over a
billion is still more than 3 MeV from the exact ground state energy,
whereas an $N_{\max}=10$ calculation with a basis dimension of about
10 million is within 1 MeV of the converged ground state energy.  Also
for other observables~\cite{Abe:2012wp}, such as radii and quadrupole
moments, the convergence seems to be slower with the FCI truncation
than with the $N_{\max}$ truncation, when viewed as function of the
basis space dimension.

Finally, with the $N_{\max}$ truncation one may easily separate the
positive parity spectrum and the negative parity spectrum.  Two
succesive shells have opposite parity, depending on whether the
orbital angular momentum is even or odd.  By increasing the total
number of quanta of the many-body basis states by two quanta at a
time, we automatically create a basis with a specific parity. The
'natural' parity of a nucleus is given by its valence space, for which
our calculations start with $N_{\max}=0$; for the 'unnatural' parity
spectrum we would perform a series of calculations starting with
$N_{\max}=1$.  In both cases, $N_{\max}$ is increased by two quanta as
we increase the basis space.  (Of course, it is straightforward to
include a parity-constraint in FCI calculations, and in fact, the
dimensions given here for FCI calculations do incorporate such a
constraint.)

\subsection{Extrapolation to the complete basis}

For a very light nucleus like $^4$He or $^6$Li one can achieve
convergence of the ground state energy and other observables by simply
going to a sufficiently large basis, but for larger nuclei that is not
practical.  However, based on a series of calculations in finite basis
spaces, we can extrapolate those results to a complete,
infinite-dimensional, basis.
Empirically~\cite{Maris:2008ax,Bogner:2007rx} the nuclear binding
energies seem to converge exponentially with $N_{\max}$
\begin{eqnarray}
  E_{\hbox{\scriptsize binding}}^{N} &=& 
    E_{\hbox{\scriptsize binding}}^{\infty} + a_1 \exp(-a_2 N_{\max}) \,.
\end{eqnarray}
There are several different ways in which one can exploit this
(empirical) convergence behavior.  Here we use three consecutive
$N_{\max}$ values at a fixed basis space parameter $\hbar\omega$ in
order to estimate a converged binding energy at that $\hbar\omega$.
Under the assumption that the convergence is indeed exponential, such
an extrapolation should get more accurate as $N_{\max}$ increases; the
difference between the extrapolated results from two consecutive sets
of three $N_{\max}$ values is then used as an estimate of the
numerical uncertainty associated with the extrapolation.  We do this
for a range of basis space $\hbar\omega$ values, in order to check the
consistency of the extrapolated results and their error estimates.
(We only use three consecutive $N_{\max}$ values for each
extrapolation, since our results at finite basis spaces are accurate
to at least five significant figures.)

\begin{figure}
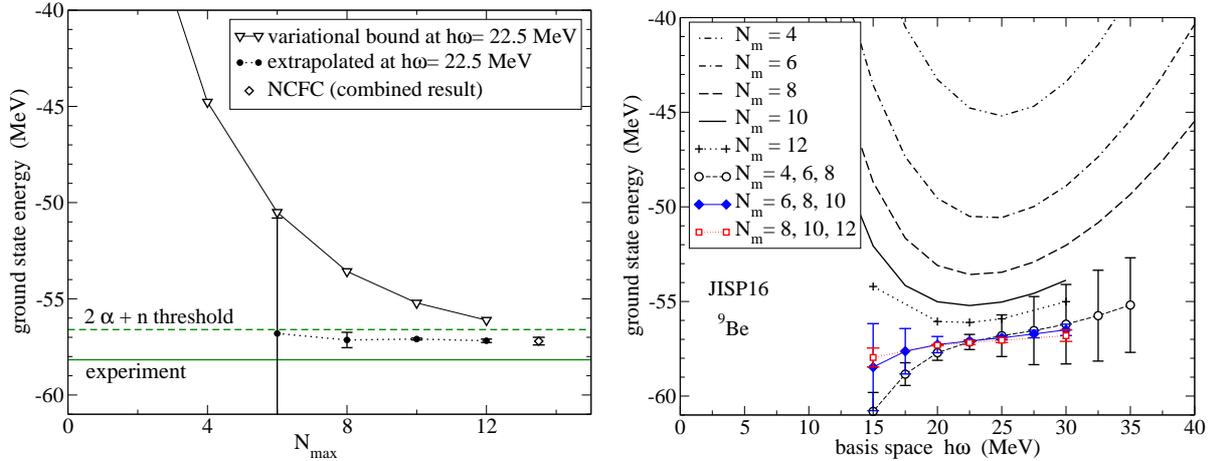

\includegraphics[width=0.48\textwidth]{results_9Be_Nm.eps}\quad
\includegraphics[width=0.48\textwidth]{results_9Be_hw.eps}
\caption{\label{Fig:9Be}
  Ground state energy of $^9$Be with JISP16 
  as function of $N_{\max}$ (left) and
  as function of $\hbar\omega$ for several $N_{\max}$ values (right),
  as well as extrapolated results with error estimates.}
\end{figure}
This procedure is illustrated for $^9$Be in Fig.~\ref{Fig:9Be}, again
with JISP16.  In the left panel we show the ground state energy for a
series of calculations at a fixed HO parameter $\hbar\omega$, as well
as the extrapolated ground state energy.  For the error estimate on
the first extrapolated value, obtained from calculations at
$N_{\max}=2$, $4$, and $6$, we use the difference between the
calculated result at $N_{\max}=6$ and the extrapolated value; for
higher $N_{\max}$ values we use the difference with the value obtained
from the next-smaller basis spaces.  

The extrapolated result appears to be very well converged
(i.e. consistent within diminishing errors) starting from
$N_{\max}=8$, not only as function of $N_{\max}$, but also as function
of the basis space parameter $\hbar\omega$, as can be seen from the
right panel of Fig.~\ref{Fig:9Be}.  For our final No-Core Full
Configuration (NCFC) result we combine the extrapolation results for a
range of HO parameters at and slightly above the variational minimum
to arrive at our best estimate for the results in the complete basis,
independent of all basis parameters.

\subsection{Need for HPC and Code performance}

In order to reliably extrapolate to the complete basis, we need
results in finite bases up to at least $N_{\max} = 8$ ($9$ for
unnatural parity states), and preferably $N_{\max} = 10$ ($11$ for
unnatural parity states) or higher.  This leads to a computational
challenge, because the basis space dimension increases rather rapidly
with $N_{\max}$ and with the number of particles.  Furthermore, the
sparsity of the matrix depends on the 'rank' of the potential: a
two-body potential leads to a much sparser many-body matrix than a
three-body potential.

\begin{figure}
\begin{minipage}{18pc}
\includegraphics[width=18pc]{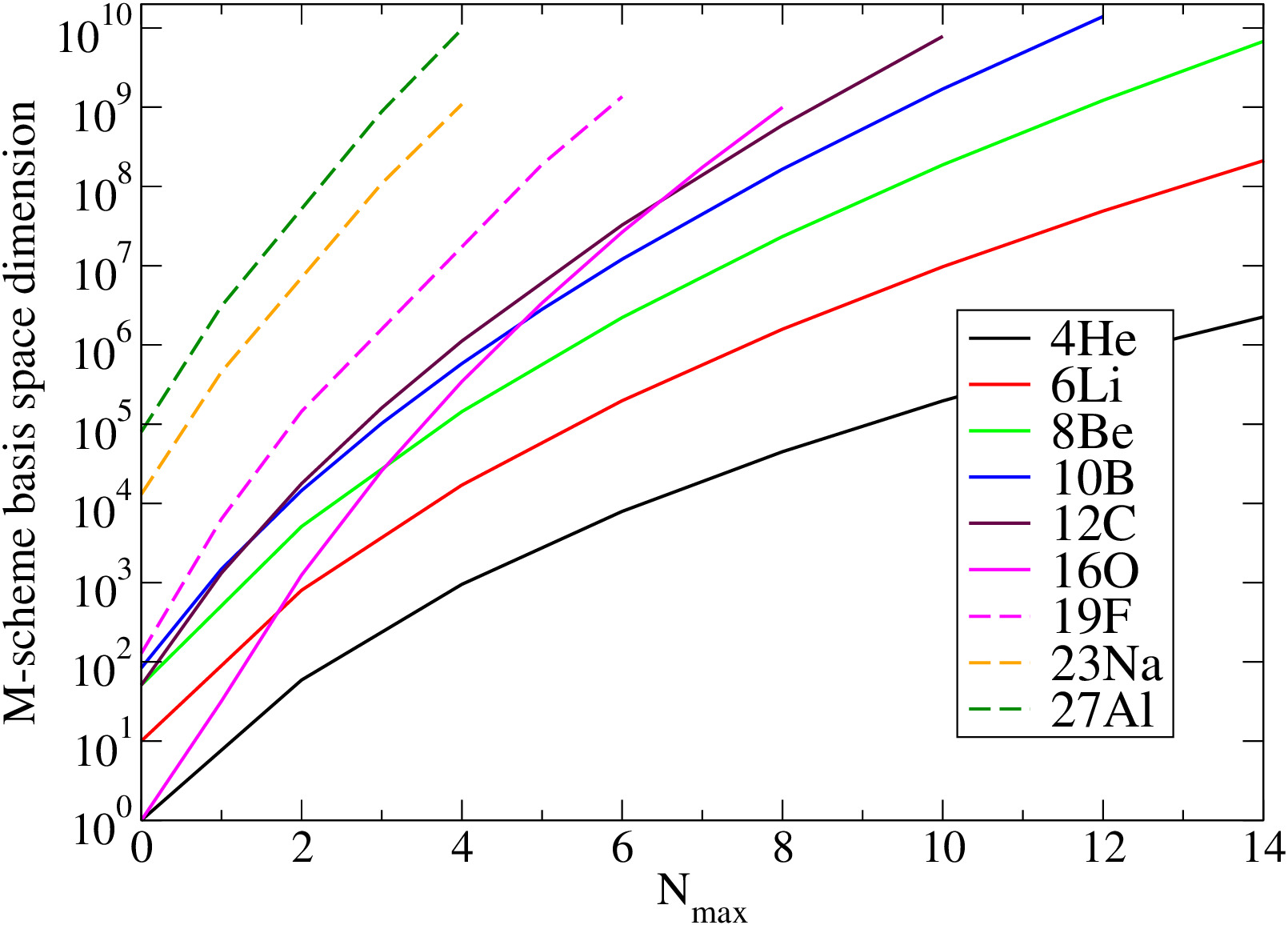}
\caption{\label{Fig:dimensions}
  Basis space dimension as function of $N_{\max}$
  for $N=Z$ nuclei in the $p$-shell (solid), 
  as well as a few nuclei in the $sd$-shell (dashed).}
\end{minipage}\hspace{2pc}%
\begin{minipage}{18pc}
\includegraphics[width=18pc]{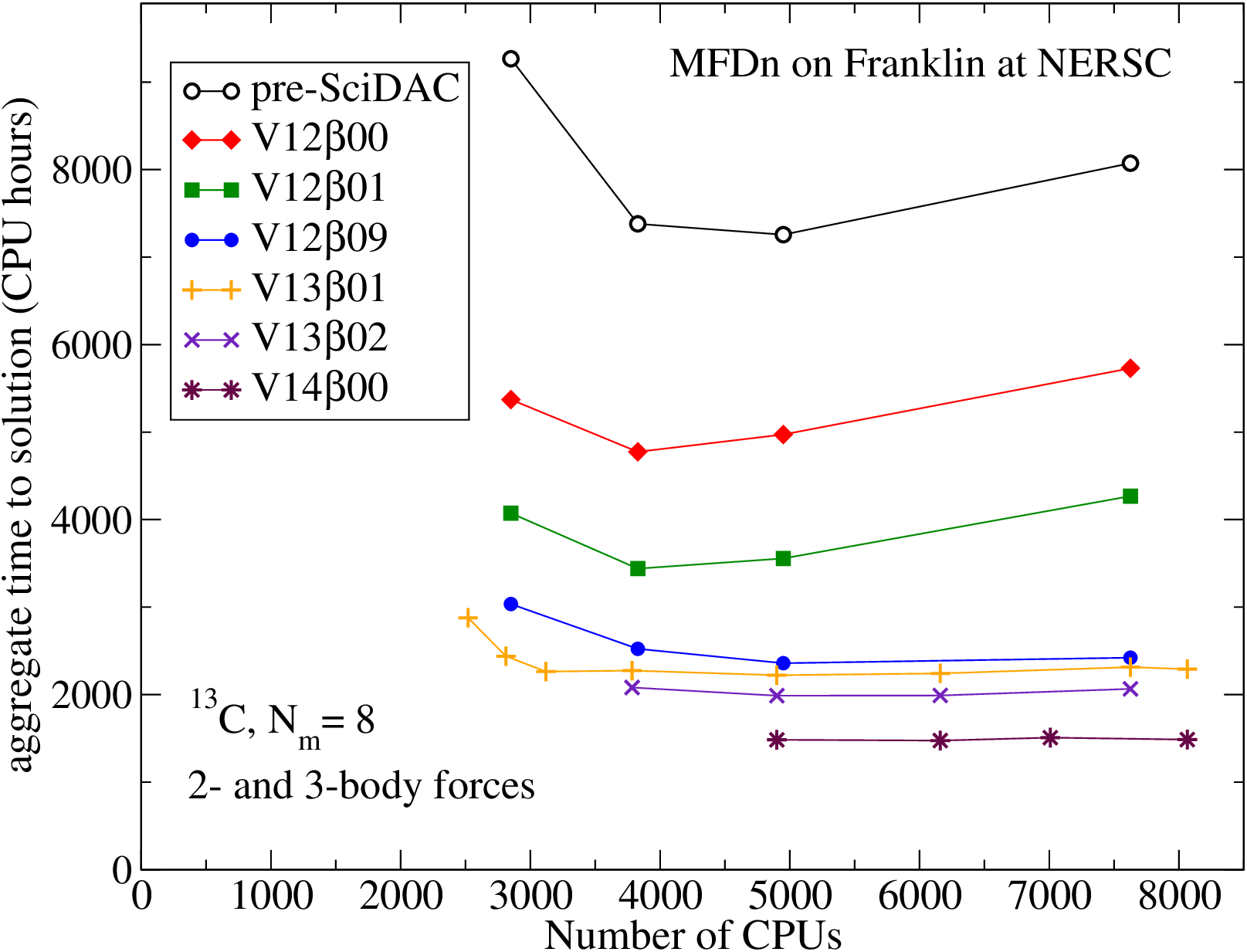}
\caption{\label{Fig:performance} 
  Performance improvements on Franklin (at NERSC) 
  of the code MFDn for $^{13}$C with 2- and 3-body forces.}
\end{minipage}
\end{figure}

In Fig.~\ref{Fig:dimensions} we show the basis space dimensions for a
number of $N=Z$ nuclei in the $p$-shell (solid curves), as well as a
few nuclei in the $sd$-shell.  The basis space dimension for even the
first couple of nuclei in the $sd$-shell (more than 8 protons and 8
neutrons) is of the order of 10 billion or more at $N_{\max} = 8$.  As
the matrix size increases, there is a clear need for high-performance
computing: obtaining the lowest 10 to 20 eigenvalues and eigenvectors
of matrices with a dimension of the order of several billion is a
nontrivial task, even if the matrix is extremely sparse.

For our calculations we use the code MFDn, which is a hybrid
MPI/OpenMP parallel Fortran 90 code for nuclear structure
calculations, that has been in development for over two decades.
Significant improvements in its performance have been made over the
last four
years~\cite{Sternberg:2008:ACI:1413370.1413386,Vary:2009qp,Maris201097}
under the UNEDF SciDAC program, as is illustrated in
Fig.~\ref{Fig:performance}.  On currently available machines, the
largest many-body basis dimension that MFDn can handle is about 10 to
15 billion with two-body forces only, and about one billion with two-
and three-body forces.

MFDn constructs the many-body basis states in $M$-scheme and the
corresponding Hamiltonian matrix, and solves for the lowest
eigenstates using an iterative Lanczos algorithm.  At the end of a
run, it writes the nuclear wavefunctions to file, and evaluates
selected physical observables which can be compared to experimental
data (the wavefunctions themselves are not experimentally observable).
It also writes out the one-body density matrix elements (in the
underlying single-particle basis), which can be used for further
analysis.

The matrix is distributed in a 2-dimensional fashion over the
processors, and we work with the lower triangle only, because the
matrix is symmetric (and real).  Because of the 2-dimensional
distribution of the matrix, it runs on $n (n+1)/2$ processors, where
$n$ is the number of ``diagonal'' processors.  Both the memory usage
and the (local) CPU usage are distributed evenly over all processors.
This load-balancing is achieved by a round-robin distribution of the
$M$-scheme basis states over the ``diagonal'' processors.  The
disadvantage of this scheme is that it destroys any natural sparsity
structure in the matrix.  We are currently working on a new
load-balancing scheme that retains some of the natural sparsity
structure (Version 14 in Fig.~\ref{Fig:performance}).  The Lanczos
vectors are also stored in memory, distributed over all processors;
this allows for fast and efficient re-orthoganalization after every
matrix-vector multiplication.  Note that the re-orthogonalization is
done in double precision, even though the matrix and vectors
themselves are stored in single precision.  This turned out to be
essential in order to maintain numerical accuracy over hundreds of
Lanczos iterations with matrix dimensions in the billions.

The code has good scaling characteristics, as illustrated by the
flatness of the curves for the more recent versions of the code in
Fig.~\ref{Fig:performance}.  For large runs on modern multi-core
architectures such as the Cray XT4/5 and XE6, it is significantly more
efficient to run the code in a hybrid mode, using OpenMP threading
within a node, and MPI between nodes~\cite{Maris201097}.  We are
currently working on improving the performance on NUMA architectures
by focussing on the data locality~\cite{Srivinasa2012256}.

The increase in aggregate CPU time as the number of CPUs goes below
4,000 in Fig.~\ref{Fig:performance} is due to the memory limitations.
One of the recent code developments was focussed on reducing the
memory footprint by partially re-generating the matrix 'on the fly'.
This enabled us to perform the largest runs with two- and three-body
forces for $^{14}$C and $^{14}$N on Jaquar at ORNL~\cite{Maris:2011as}.

\section{Emergence of clustering and rotational phenomena in Beryllium isotopes}

Among the Be-isotopes there are several interesting and challenging
problems for ab initio nuclear structure calculations:
$\alpha$-clustering is likely to play an important role, in particular
for $^8$Be and $^9$Be; there are several rotational bands; and they
are among the lightest nuclei for which both positive and negative
parity states are known experimentally.  With the NCFC approach
described above, we should be able to perform reasonably accurate
calculations for the Be-isotopes from $^6$Be up to $^{12}$Be, provided
we use a suitably soft interaction such as JISP16.

\subsection{Clustering}

We have already presented our results for the ground state energy of
$^9$Be in Fig.~\ref{Fig:9Be}.  With JISP16 we find a ground energy of
$E_{\rm gs}= -57.2(2)$~MeV, compared to an experimental value of
$E_{\rm gs}= -58.167$~MeV, that is JISP16 underbinds $^9$Be by about
$1$~MeV.  Despite this underbinding, our result is below the threshold
for two $\alpha$ particles (plus a neutron), which is at $-56.6$~MeV.
Thus we are dealing with a true bound state, not a resonance state.

The next question that naturally arises is: what is the structure of
this state?  In particular, does it have a structure of two $\alpha$
particles plus a neutron, as one might expect?  In an attempt to
answer such questions, we have looked at the proton and neutron
densities in coordinate-space.

The local density for a nucleus with wavefunction $\Psi$ is given by
\begin{eqnarray}
 \rho(\vec{r}) 
  &=& \int \Psi^\star(\vec{r}, \vec{r}_2, \ldots, \vec{r}_A) \; 
           \Psi(\vec{r}, \vec{r}_2, \ldots, \vec{r}_A) \;
           d^3r_2 \ldots d^3r_A \,.
\label{Eq:localdens}
\end{eqnarray}
This corresponds to the probability of finding a nucleon at position
$\vec{r}$.  Since we use a basis in single-particle coordinates,
rather than relative coordinates, our wavefunction $\Psi(\vec r_i)$
includes cm motion.  The local density extracted from such
wavefunctions therefore includes also contributions from the cm
motion.  However, because of the exact factorization of the cm
wavefunction, see Eq.~(\ref{Eq:exactfact}), this density is a
convolution of the cm density and a
translationally-invariant density $\rho_{\rm ti}$ relative to the cm
of the entire nucleus
\begin{eqnarray}
  \rho^\omega(\vec{r}) &=&
     \int \rho_{\rm ti}(\vec{r}-\vec{R}) \; 
     \rho^\omega_{\rm cm}(\vec{R})\; d^3\vec{R} \,.
\end{eqnarray}
The cm density $\rho^\omega_{\rm cm}$ is a simple Gaussian that smears
out $\rho_{\rm ti}$ and obfuscates details of the local
density\footnote{Note that the cm motion also introduces a spurious
  dependence on the basis parameter $\hbar\omega$ into $\rho^\omega$
  that masks the convergence.  Even in the limit of a converged
  calculation in a complete basis, $\rho^\omega$ depends on
  $\hbar\omega$, whereas the translationally-invariant density
  $\rho_{\rm ti}$ becomes independent of the basis as one approaches
  convergence.}.  
The translationally-invariant density $\rho_{\rm ti}$ can be obtained
by a deconvolution of the cm density using standard Fourier
methods~\cite{cockrell2012}
\begin{eqnarray}
 \rho_{\rm ti}(\vec{r})&=& F^{-1}\bigg[
   \frac{F[\rho^\omega(\vec{r})]}{F[\rho^\omega_{\rm cm}(\vec{r})]}\bigg] \,,
\end{eqnarray}
where $F[f(\vec r)]$ is the 3D Fourier transform of $f(\vec r)$.  Note
that the densities on the RHS of this equation depend on the basis
space parameter $\hbar\omega$, even in the infinite-basis limit, but
the LHS is independent of the basis in this limit.  

\begin{figure}
\includegraphics[width=0.32\textwidth]{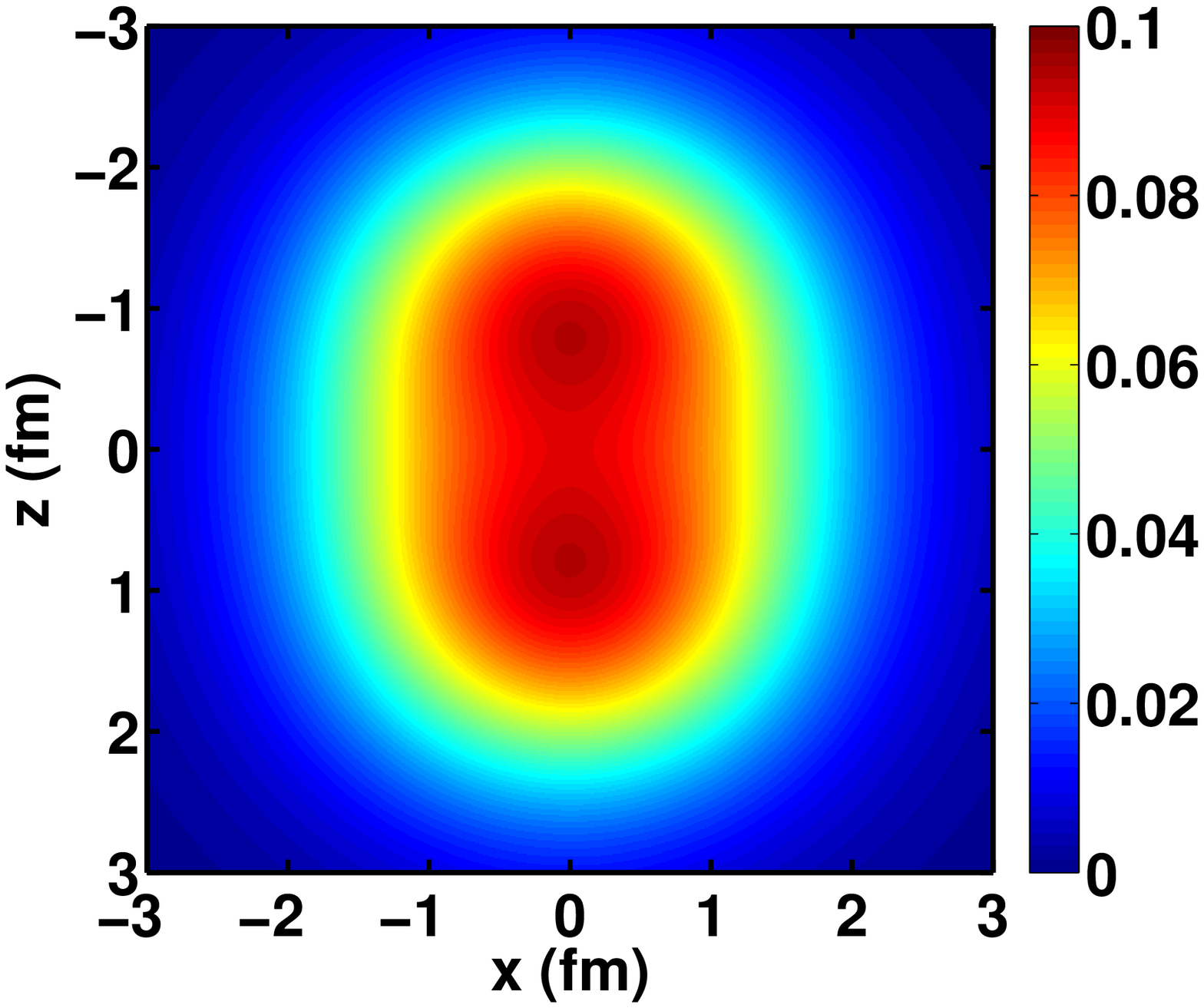}\quad
\includegraphics[width=0.32\textwidth]{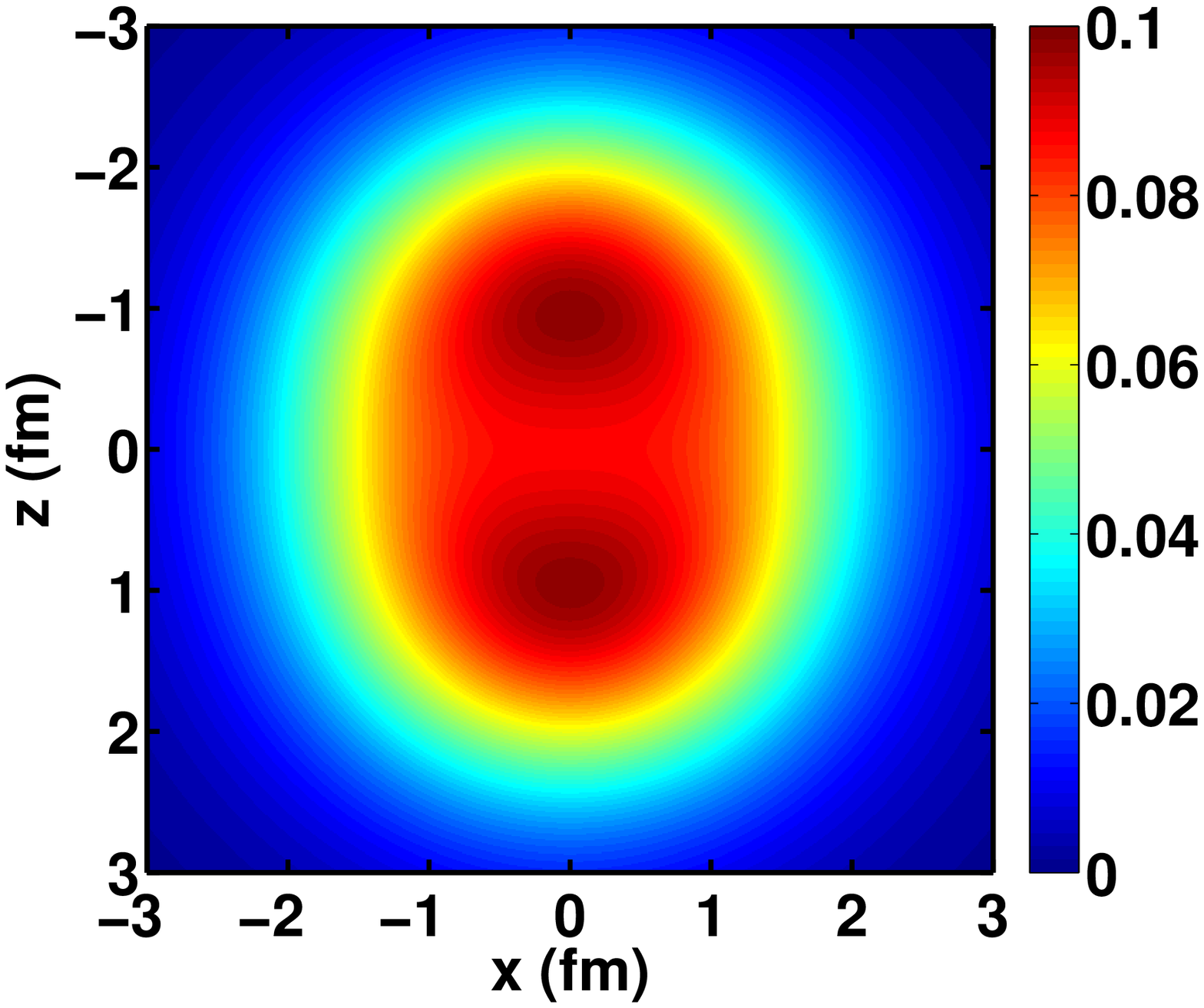}\quad
\includegraphics[width=0.32\textwidth]{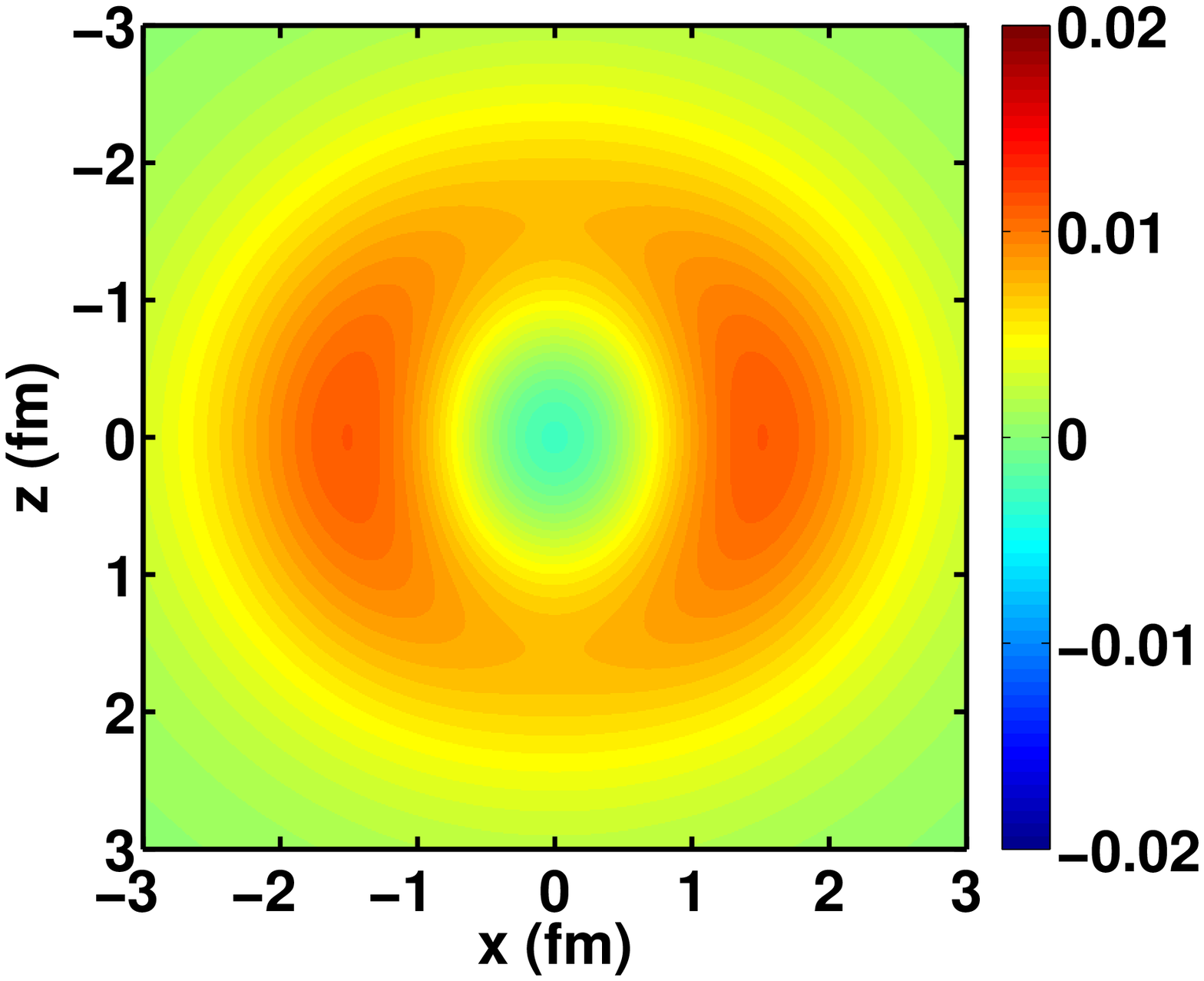}
\caption{\label{Fig:9Be_dens}
  Proton (left), neutron (center), and the difference between the
  neutron and proton density (right) for the $J^\pi=\frac{3}{2}^-$
  ground state of $^9$Be, with $M_j=\frac{3}{2}$ using JISP16.  }
\end{figure}
In Fig.~\ref{Fig:9Be_dens} we show the proton and neutron densities
for the $J^\pi=\frac{3}{2}^-$ ground state of $^9$Be, with
$M_j=\frac{3}{2}$, obtained with JISP16 in a $N_{\max}=10$,
$\hbar\omega=12.5$~MeV basis space, where the RMS radii seem to be
most converged.  The spurious cm contribution has been removed as
described above producing the translationally-invariant neutron and
proton densities shown in Fig.~\ref{Fig:9Be_dens}.  The figure shows
the density in the $xz$-plane, and is symmetric along the $z$-axis
(vertical axis).  It clearly shows a clustering of both the protons
and the neutrons into two compact dense regions, seperated by about 2
fm along the $z$-axis.  Although this does not prove that we have
$\alpha$-clustering in our ab initio calculation, it is at least
consistent with $\alpha$-clustering.  Furthermore, the difference of
the neutron and proton density (right panel of
Fig.~\ref{Fig:9Be_dens}) shows that the 'extra' neutron forms a ring
or donut shape around the $z$-axis, rather than being centered in
between the dominant proton and neutron clusters.

\subsection{Rotation}

Rotational bands arise quite naturally in $\alpha$-cluster models for
the Be isotopes~\cite{rotbands}, so now that we do see some
evidence for clustering in $^9$Be, we can also expect the
corresponding rotational states in the spectrum.  The rotational
energy for states in a rotational band with an axial symmetry is given
by~\cite{textbook}
\begin{eqnarray}
 E_{\hbox{\scriptsize{rotational}}} &=& 
           \frac{\hbar^2}{2 {\cal I}} \; \left( J (J+1) - K^2 \right) \,,
\label{Eq:Energy}
\end{eqnarray}
where ${\cal I}$ is the moment of inertia, $K$ is the projection of $J$
along the intrinsic symmetry axis, and $J$ is the total angular momentum.
Thus rotational bands show up as straight lines if we plot the
excitation energies as function of $J(J+1)$.  

\begin{figure}
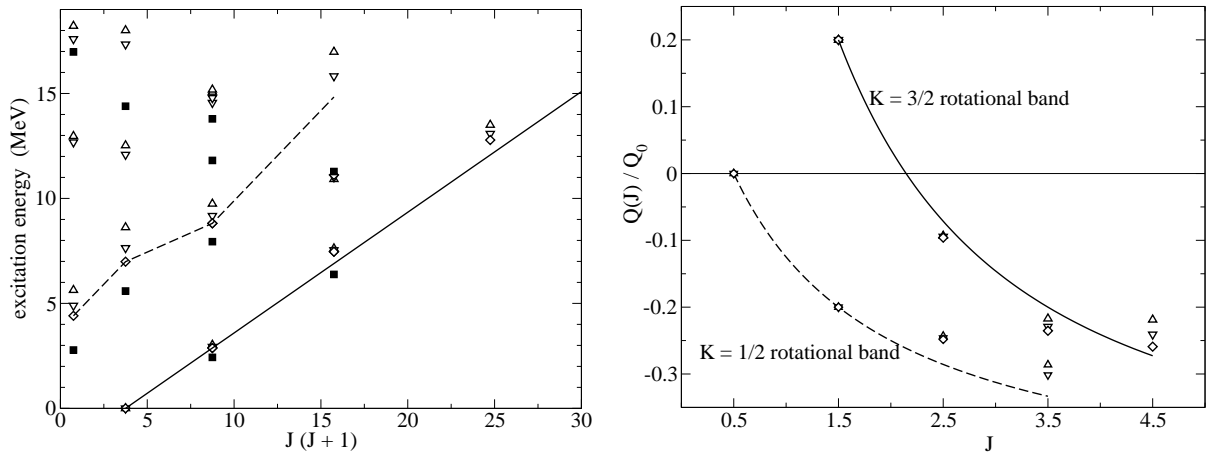

\includegraphics[width=0.48\textwidth]{rotational_9Be_nat_E.eps}\quad
\includegraphics[width=0.48\textwidth]{rotational_9Be_nat_Q.eps}
\caption{\label{Fig:9Be_rot}
  Evidence for the ground state $K=3/2$ (\full)
  and excited $K=1/2$ (\dashed) rotational bands in $^9$Be 
  from ab initio calculations with JISP16 (\opentriangle : $N_{\max}=8$,
  \opentriangledown : $N_{\max}=10$, \opendiamond : $N_{\max}=12$)
  Left: Negative parity spectrum as function of $J(J+1)$
  (\fullsquare : experimental data from Ref.~\cite{Tilley2004155}.);
  Right: Quadrupole moments of the rotational bands 
  normalized by $Q_0$.}
\end{figure}

For the ground state rotational band in $^9$Be we have
$K=J=\frac{3}{2}$.  Indeed, the spectrum shown in the left panel of
Fig.~\ref{Fig:9Be_rot} suggests that the lowest $J=\frac{5}{2}$,
$J=\frac{7}{2}$, and $J=\frac{9}{2}$ states form a rotational band
with the ground state.  Furthermore, the excitation energy of the
members of this rotational band is very well converged, in particular
for the $J=\frac{5}{2}$ and $J=\frac{7}{2}$ states.  The calculated
excitation energy also agrees quite well with the experimental
spectrum, though we do predict a $J=\frac{9}{2}$ member of this
rotational band, in addition to the confirmed experimental
$J=\frac{5}{2}$ and $J=\frac{7}{2}$ states.

The spectrum by itself is not sufficient to claim that these states
are indeed rotational states.  The rotational model also predicts the
quadrupole moments of a rotational band to all be related to an
'intrinsic' quadrupole moment $Q_0$~\cite{textbook}
\begin{eqnarray}
 Q(J)   &=& \frac{3 K^2 - J (J+1)}{(J+1) (2J+3)} \; Q_0 \,.
\label{Eq:Quad}
\end{eqnarray}
The strength of the B(E2) transitions within a rotational band also
follow from this intrinsic quadrupole moment $Q_0$.  Even though
the quadrupole moments are not very well converged in our
calculations\footnote{The HO basis functions fall off like
  $\exp(-r^2)$ whereas the asymptotic behavior of the nuclear
  wavefunction is known to be an expontial, $\exp(-r)$, rather than a
  Gaussian. The quadrupole moments are sensitive to this long-range
  part of the nuclear wavefunction that is not very well represented
  in a finite HO basis.}, the {\em ratio} of the quadrupole moments
does seem to be reasonably well converged.  In the right panel of
Fig.~\ref{Fig:9Be_rot} we show the quadrupole moments, normalized by
the intrinsic quadrupole moment, $Q_0$, extracted from the
ground state.  They agree quite well with the prediction of
Eq.~(\ref{Eq:Quad}).  Also the calculated B(E2) transitions are in
reasonable agreement with the rotational model~\cite{caprio2012}.

The other states in the low-lying negative-parity spectrum of $^9$Be
are not as well converged as the ground state rotational band, as can
be seen from the left panel of Fig.~\ref{Fig:9Be_rot}.  However, our
calculations suggests that there is also a $K=\frac{1}{2}$ rotational
band, starting from the first $J=\frac{1}{2}$ state.  Note that it
involves the third $J=\frac{7}{2}$ state, rather than the second.  Not
only are the excitation energies in agreement with predictions from a
rotational model (including the staggering due to the Coriolis
effect), the quadrupole moments also agree with Eq.~(\ref{Eq:Quad}).
In addition, the convergence pattern of these four states is quite
similar.  The lowest three states of this band seem to converge
towards the experimental excitation energies.  More work is needed to
determine whether these states indeed form a rotational
band~\cite{caprio2012}.

\subsection{Ground state energies and parity splittings}

\begin{figure}
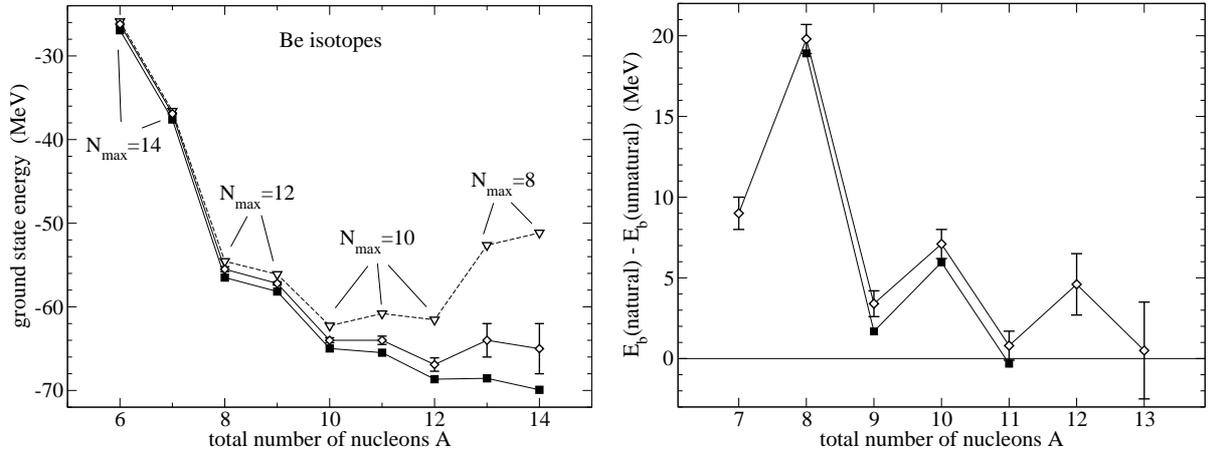

\includegraphics[width=0.48\textwidth]{Eground.eps}\quad
\includegraphics[width=0.48\textwidth]{DeltaE_unn_nat.eps}
\caption{\label{Fig:Beground}
  Ground state energies (left) and parity splittings (right) 
  of Be-isotopes with JISP16 
  (variational upperbound: \opentriangledown, NCFC: \opendiamond),
  together with experimental data (\fullsquare) from
  Refs.~\cite{Tilley2004155,Audi19971}.}
\end{figure}

With JISP16 we have not only looked at $^9$Be, but also at the other
Be-isotopes, from $^6$Be up to $^{14}$Be.  Because of the variational
principle, any result for the ground state energy in a finite basis
truncation forms a strict upper bound for the exact ground state
energy.  Indeed, our calculated energies in finite basis spaces tend
to be well above the experimental ground state energies, see
Fig.~\ref{Fig:Beground}.  With an exponential extrapolation on the
total binding energies we obtain results for the ground state energies
(with numerical error bars due to the extrapolation uncertainties)
that are in general much closer to the experimental ground state
energies than these variational upper bounds.  It turns out that with
JISP16 all Be-isotopes are underbound (left panel of
Fig.~\ref{Fig:Beground}): the light isotopes, $^6$Be and $^7$Be, by a
fraction of an MeV, and the neutron-rich isotopes $^{13}$Be and
$^{14}$Be, by about $5$ MeV.  Nevertheless, the overall pattern of the
ground state energies is in quite good agreement with the data.  Also
the spectra and several other observable quantities do seem to be in
quite good agreement with the data, despite this overall underbinding.

Starting with $^8$Be, states of both partities appear in the low-lying
spectrum; and for $^{11}$Be the lowest positive parity state is the
ground state, contrary to the expectations based on the shell model,
which predicts negative parity ground states for all odd $p$-shell
nuclei.  For $^7$Be through $^{13}$Be we performed our calculations
for both natural parity states and for unnatural parity states.  We
then perform independent extrapolations for the lowest positive parity
states and the for the lowest negative parity states.

In the right panel of Fig.~\ref{Fig:Beground} we show the difference
between the binding energy of the lowest natural parity state and the
lowest unnatural parity state.  For most isotopes we expect this
difference to be positive, but for isotopes with parity inversion it
becomes negative.  Although our results do not quite reproduce the
observed parity inversion for $^{11}$Be, parity inversion is within
our numerical error estimates for this isotope.  Furthermore, over the
range of isotopes from $^{8}$Be to $^{11}$Be our results are in very
good qualitative agreement with the data: with JISP16 we seem to
underbind all unnatural parity states by a similar amount of about
$1$~MeV.  Based on these results, we also predict parity inversion for
$^{13}$Be; experimentally, the parity of the ground state is not
confirmed, though likely to be negative~\cite{Audi19971}, which indeed
implies parity inversion ($^{13}$Be has one neutron in the $sd$-shell,
so the natural parity is positive).

It is not yet clear whether the relative underbinding of the unnatural
parity states compared to the natural parity states in beryllium is a
deficiency of the NN interaction, JISP16, or due to underestimating
the numerical error in the extrapolation to the infinite basis, in
particular for the unnatural parity states~\cite{Forssen:2004dk}.
Also the general underbinding of the ground states hints at a
deficiency of JISP16.  Indeed, it is unlikely that this interaction is
'perfect': it was fitted to the NN scattering data, as well as to
select observables in light nuclei, in particular to $^6$Li.
Nevertheless, the overall agreement between data and our calculation
is quite encouraging, and we expect to be able to get even closer to
the data with improved NN and 3NF potentials.

For a true ab initio calculation one would obtain the nuclear forces
(NN, 3NF, etc.) from the underlying quantum field theory, QCD.
Unfortunately, that is not yet possible without introducing free
parameters.  Based on QCD, one can derive nonrelativistic effective
NN, 3NF, and even 4NF nuclear potentials using chiral perturbation theory
(ChPT)~\cite{Entem:2003ft,Epelbaum:2005pn}.  Such a derivation
determines the structure of the nuclear interaction, but not the
numerical constants that govern the strength of the different terms in
the interaction -- these have to be fitted to low-energy nuclear
physics data.  (In the future they may be calculable through lattice
QCD.)

\section{Understanding the Gamow--Teller transitions between $^{14}$C and $^{14}$N}

The anomalously long lifetime of $^{14}$C, $5730 \pm 30$ years,
compared to lifetimes of other light nuclei undergoing the same decay
process, allowed Gamow--Teller (GT) beta-decay, poses a major
challenge to ab initio nuclear structure calculations.  Not only is
the number of particles (14) pushing the limit of what ab initio
calculations can reliably calculate, the experimental lifetime can
only be reproduced by an anomalously small matrix element for this
transition.  This suggests that delicate cancellations could play a
crucial role, which always poses a challenge to numerical
calculations.  On the other hand, since the transition operator, in
leading approximation, depends only on the nucleon spin and isospin
but not on the spatial coordinate, this decay provides a precision
tool to inspect selected features of the initial and final states.

Traditional realistic NN forces alone appear insufficient to produce
the observed lifetime~\cite{Aroua:2002xm,Negret:2006zz}, indicating
that three-body forces may be needed.  Indeed, in
Ref.~\cite{Maris:2011as} we showed that the 3NF of ChPT plays a major
role in producing a transition rate that is near zero, needed for the
anomalous long lifetime.  We used the chiral two- and three-body
potentials of ChPT~\cite{Entem:2003ft,Epelbaum:2005pn} and the No-Core
Shell Model (NCSM) using Lee--Suzuki (LS) renormalization~\cite{LSO}
to improve the convergence.  The non-perturbative coupling constants
of the 3-body force, not fixed by $\pi-N$ or NN data, are $c_D$ for
the $N-\pi-NN$ contact term and $c_E$ for the $3NF$ contact term.  In
Ref.~\cite{Maris:2011as} we presented results for
$(c_D,c_E)=(-0.2,-0.205)$ and for $(-2.0,-0.501)$.  Both sets fit the
A=3 binding energies and are allowed by the 'naturalness criterium'
for these parameters.  The former also produces a precise fit to the
triton half life~\cite{Gazit:2008ma}; the latter produces the triton
half life within $20 \%$ of experiment but is preferred by the
$^{14}$C half life.

\begin{figure}[b]
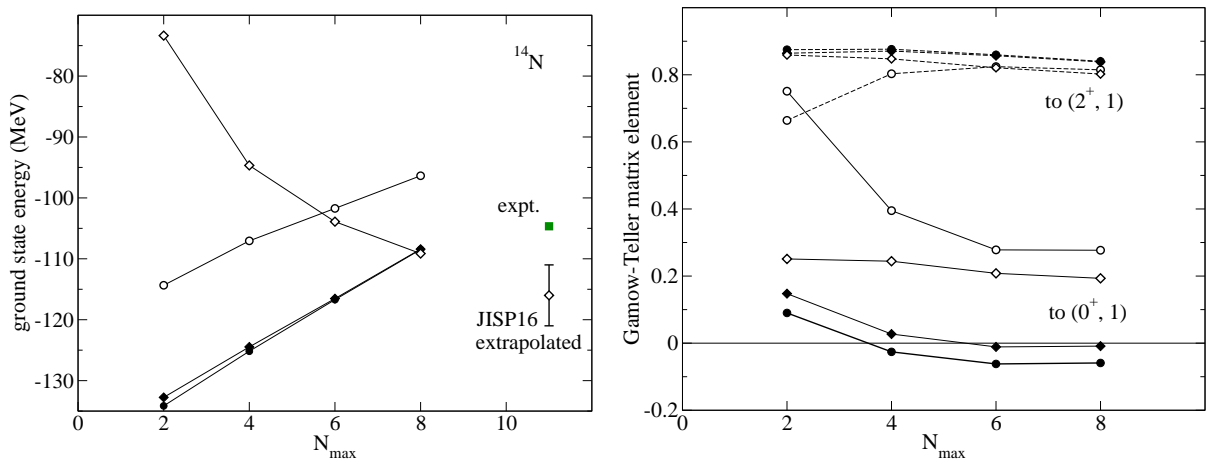

\includegraphics[width=0.48\textwidth]{res_Nm_Egs.eps}\quad
\includegraphics[width=0.48\textwidth]{res_Nm_MGT.eps}
\caption{\label{Fig:14N}
  Ground state energies of $^{14}$N (left) and GT matrix element (right) 
  as function of $N_{\max}$ with JISP16 at $\hbar\omega = 27.5$~MeV (\opendiamond),
  with chiral two-body forces at $\hbar\omega = 14.0$~MeV (\opencircle),
  with chiral two- and three-body forces at $\hbar\omega = 14.0$~MeV
  with $(c_D,c_E)=(-0.2,-0.205)$ (\fullcircle) and
  with $(c_D,c_E)=(-2.0,-0.501)$ (\fulldiamond), 
  and experimental data (\fullsquare).}
\end{figure}
In the left panel of Fig.~\ref{Fig:14N} we show the obtained binding
energy of $^{14}$N both with the chiral two- and three-body forces and
with the phenomenological two-body interaction JISP16.  For JISP16 we
can use our extrapolation technique to extract the binding energy in
the complete basis space: we find 116(5)~MeV, that is, $^{14}$N is
overbound by about 10~MeV with JISP16.  For the chiral two- and
three-body forces we used a LS renormalization procedure on the
truncated basis space, which means that the results are no longer
variational upperbounds, and the convergence to the infinite basis
space is not monotonic.  Our results for the binding energy do not
imply convergence with $N_{\max}$; however, the obtained spectrum at
$N_{\max}=8$ appears to be reasonably well
converged~\cite{Maris:2011as} and in qualitative agreement with the
experimental spectrum, both for $^{14}$C and for $^{14}$N.
Furthermore, in general the LS ground state energies increase with
$N_{\max}$ at small and moderate $N_{\max}$ values, but this trend
tends to turn around at (very) large $N_{\max}$ values, and in the
limit $N_{\max}\to\infty$ the ground state energy tends to converge
from above, as is the case without LS renormalization.

The GT matrix element that governs the $\beta$-decay of $^{14}$C is
\begin{eqnarray}
M_{\rm GT}  &=& \sum_{\alpha,\beta}  
\langle\beta|\sigma\tau_+|\alpha\rangle \; \rho_{\alpha\beta} \,,
\label{Eq:MGT}
\end{eqnarray}
where $\langle \alpha|\sigma\tau_+|\beta \rangle$ is the one-body matrix
element between HO single-particle states $\alpha$ and $\beta$, 
which is non-vanishing only when both single-particle states are in the same shell, 
and the one-body density matrix 
\begin{eqnarray}
  \rho_{\alpha\beta} &\equiv& 
  \langle \Psi_f|a_{\beta}^{\dagger}a_{\alpha}|\Psi_i\rangle \,,
\end{eqnarray}
with $\langle\Psi_f|$ and $|\Psi_i\rangle$ the $^{14}$N and $^{14}$C
ground states respectively.  For comparison, we also calculate the GT
transition between the first $2^+$ state in $^{14}$C and the ground
state in $^{14}$N.  In order to reproduce the measured half life of
$T_{1/2} \simeq 5730$ years, the GT matrix element must be anomalously
small, $|M_{\rm GT}^{M}| \simeq 2 \times 10^{-3}$, in contrast with a
conventional strong GT transition between states in light nuclei with
$|M_{\rm GT}| \simeq 1$.  

In the right panel of Fig.~\ref{Fig:14N} we show GT matrix
elements\footnote{For simplicity, these results are obtained by
  calculations by assuming isospin symmetry in $^{14}$N, and
  considering transitions between the $(1^+,0)$ ground state and the
  lowest $(0^+,1)$ and $(2^+,1)$ excited states in $^{14}$N.}  both
with the chiral two- and three-body forces and with JISP16.  The
results for the GT matrix element between the ground state of $^{14}$N
and the lowest $(J^\pi,T)=(2^+,1)$ excited state of $^{14}$C is
reasonably well converged for the different interactions.
Furthermore, it is insensitive to the interaction: JISP16 gives
approximately the same result as the chiral interaction, with or
without 3-body forces.  Our calculations are in reasonable agreement
with experiment: the measured transition rate from the ground state in
$^{14}$N to the low-lying $2^+$ states in $^{14}$C is $\sum B(GT) =
0.92(33)$~\cite{Negret:2006zz}, compared to our calculated value for
the transition rate of about 1.6.

The situation is very different for the GT matrix element between the
ground state of $^{14}$N and the $(0^+, 1)$ ground state of $^{14}$C.
The results with JISP16 do not depend much on the basis space
truncation, and the matrix element is suppressed compared to a
'natural' value of the order of one, but not sufficiently suppressed
in order to explain the long lifetime of $^{14}$C.  The results with
the chiral NN forces alone depend much stronger on $N_{\max}$, but
they do seem to be converged at $N_{\max}=8$, also at a value that is
significantly smaller than the 'natural' value of one, but not
sufficiently suppressed in order to explain the long lifetime of
$^{14}$C.  However, once we add the chiral 3NF, we do get a further
suppression of the GT matrix element to almost zero, depending on the
exact values of $(c_D,c_E)$.

\begin{figure}
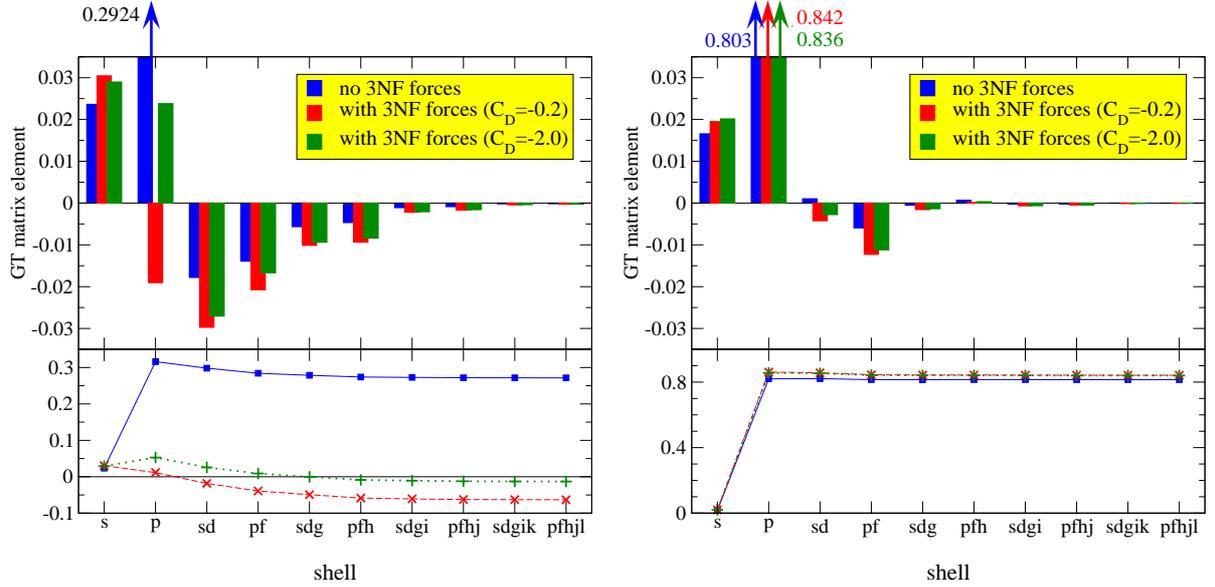

\includegraphics[width=0.48\textwidth]{MGT_14C0to14N_shells_sum_CD0.2+CD2.0.eps}\quad
\includegraphics[width=0.48\textwidth]{MGT_14C2to14N_shells_sum_CD0.2+CD2.0.eps}
\caption{\label{Fig:14Ndetails}
  GT matrix element for the transitions $^{14}$N to the $0^+$ (left) 
  and $2^+$ (right) states of $^{14}$C, decomposed per HO shell, 
  obtained with two-body chiral interaction with (red and green)
  and without (blue) three-body forces (3NF) at $N_{\max}=8$.
  Top panels display the contributions summed within the shell 
  to yield a total for that shell.  
  Bottom panels display the running sum of the GT contributions 
  over the shells included in the sum.
  Note that the top panels have the same scale left and right,
  but the bottom panels do not.}
\end{figure}
In order to reveal what drives this suppression, we show in
Fig.~\ref{Fig:14Ndetails} the decomposition of $M_{\rm GT}$ at
$N_{\max}=8$ into the contributions arising from each HO shell for two
cases with the 3NF ($c_D = -0.2, -2.0$) and one without.  On the left
we show our results for the transition between the ground states of
$^{14}$N and $^{14}$C (corresponding to $\beta$-decay of $^{14}$C); on
the right the transition between the ground state of $^{14}$N and the
first excited $2^+$ state of $^{14}$C.  The largest effect occurs in
the $p$-shell for the transition between the ground states of $^{14}$N
and $^{14}$C, where the 3NF reduces the contributions by an
order-of-magnitude from the result with the NN interactions only.
This happens for both values of $c_D$; also note that the transition
between the ground state of $^{14}$N and the first excited $2^+$ state
of $^{14}$C receives its largest contribution from the $p$-shell, and
is barely affected by the 3NF.  In addition, the three-body forces
significantly enhance the contributions of each of the higher shells,
for both transitions.  In combination with the strong cancellations
within the $p$-shell for the GT matrix element between the ground
states of $^{14}$N and $^{14}$C, the contributions of the higher
shells overwhelm that of the $p$-shell, whereas these higher shells
contribute almost nothing to the value of the GT matrix element
between the ground state of $^{14}$N and the first excited $2^+$ state
of $^{14}$C.

\begin{figure}[b]
\begin{minipage}{21pc}
\includegraphics[width=20pc]{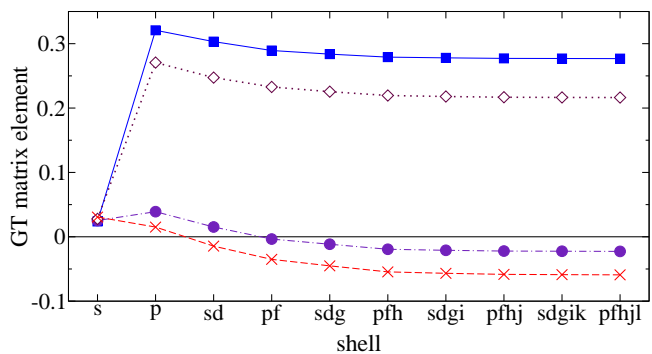}
\end{minipage}\hspace{2pc}%
\begin{minipage}{15pc}
\caption{\label{Fig:14Nmixed}
  $M_{GT}$ between the ground states of $^{14}$N and $^{14}$C,
  using the $^{14}$N wavefunction obtained with 3NF, but the $^{14}$C
  wavefunction obtained without 3NF (purple \fullcircle), and vice
  versa (maroon \opendiamond).  For comparison, we also include the
  results with (red cross) and without (blue \fullsquare) 3NF 
  for both wavefunctions.}
\end{minipage}
\end{figure}
Finally, we calculated the GT matrix element using one wavefunction
obtained with three-body forces, but the other wavefunction obtained
without three-body forces.  Fig.~\ref{Fig:14Nmixed} clearly shows that
it is the $(1^+,0)$ ground state of $^{14}$N, rather than the
$(0^+,1)$ ground state of $^{14}$C, that causes the cancellations
within the $p$-shell once the three-body forces are included.  This
seems to be a recurring theme of three-body forces in the $p$-shell
nuclei -- they manifest themselves in particular in the odd-odd
nuclei~\cite{Navratil:2007we}.

Of course, the precise value of the strongly suppressed GT matrix
element between the ground states of $^{14}$N and $^{14}$C depends on
details of the interaction, of the operator, and of the calculations,
because of the cancellations between the dominant contributions.
Although meson-exchange currents do not contribute significantly to GT
transition with a 'natural' value of the order of
one~\cite{Vaintraub:2009mm}, they may be important for the actual
value of anomalously long lifetime of $^{14}$C.  For a purely
phenomenological interaction like JISP16 it is unclear what the
appropriate corrections to the canonical operator would be; but with
the chiral interactions one should use consistent meson-exchange
currents, and apply the same renormalizaton procedure to the operators
as to the interaction.  These effects are currently under
investigation.
  
\ack 

We would like to thank Takashi Abe, Mark Caprio, Chase Cockrell, and
James Vary for useful discussions.  This work was supported in part by
U.S. Department of Energy Grant DE-FC02-09ER41582 (SciDAC/UNEDF) and
DE-FG02-87ER40371, and by the U.S. NSF grant 0904782.  Computational
resources were provided by the National Energy Research Supercomputer
Center (NERSC), which is supported by the DOE Office of Science, and
by an INCITE award, "Nuclear Structure and Nuclear Reactions", from
the DOE Office of Advanced Scientific Computing.  This research used
resources of the Oak Ridge Leadership Computing Facility at ORNL,
which is supported by the DOE Office of Science under Contract
DE-AC05-00OR22725.

\section*{References}
%

\begin{thebibliography}{10}
\expandafter\ifx\csname url\endcsname\relax
  \def\url#1{{\tt #1}}\fi
\expandafter\ifx\csname urlprefix\endcsname\relax\def\urlprefix{URL }\fi
\providecommand{\eprint}[2][]{\url{#2}}

\bibitem{Wiringa:1994wb}
Wiringa R~B, Stoks V~G~J and Schiavilla R 1995 {\em Phys. Rev.\/} {\bf C51} 38 

\bibitem{Entem:2003ft}
Entem D~R and Machleidt R 2003 {\em Phys. Rev.\/} {\bf C68} 041001

\bibitem{Pieper:2001ap}
Pieper S~C, Pandharipande V~R, Wiringa R~B and Carlson J 2001 {\em Phys. Rev.\/} {\bf C64} 014001

\bibitem{Epelbaum:2005pn}
Epelbaum E 2006 {\em Prog. Part. Nucl. Phys.\/} {\bf 57} 654--741

\bibitem{Navratil:2007we}
Navratil P, Gueorguiev V~G, Vary J~P, Ormand W~E and Nogga A 2007 {\em Phys. Rev. Lett.\/} {\bf 99} 042501 

\bibitem{bertsch2007}
Bertsch G F, Dean D J and Nazarewicz W 2007 {\em SciDAC Review} {\bf 6} 42

\bibitem{furnstahl2011}
Furnstahl R J (for the UNEDF Council) 2011 {\em Nuclear Physics News} {\bf 21} 18

\bibitem{nam2011}
Nam H {\em et~al.\/} 2012 {\em these proceedings} 
  (\textit{Preprint} \eprint{1205.227})

\bibitem{cockrell2012}
Cockrell R, Maris P and Vary J~P 2012 {\em Phys. Rev.\/} {\bf C86} 034325
  (\textit{Preprint} \eprint{1201.0724})

\bibitem{caprio2012CS}
Caprio M~A, Maris P and Vary J~P 2012 {\em Phys. Rev.\/} {\bf C86} 034312
  (\textit{Preprint} \eprint{1208.4156})

\bibitem{Gloeckner1974313}
Gloeckner D and Lawson R 1974 {\em Phys. Lett.\/} {\bf B53} 313

\bibitem{Abe:2011pk}
Abe T, Maris P, Ostuka T, Shimizu N, Ustuno Y and Vary J
2011 {\em AIP Conf. Proc.\/} {\bf 1355} 173

\bibitem{Abe:2012wp} 
Abe T, Maris P, Otsuka T, Shimizu N, Utsuno Y and Vary J 2012 
  {\em Phys. Rev.\/} {\bf C} in press
  (\textit{Preprint}~\eprint{1204.1755})

\bibitem{Shirokov:2005bk}
Shirokov A~M, Vary J~P, Mazur A~I and Weber T~A 2007 {\em Phys. Lett.\/} {\bf B644} 33 

\bibitem{Maris:2008ax}
Maris P, Vary J~P and Shirokov A~M 2009 {\em Phys. Rev.\/} {\bf C79} 014308

\bibitem{Bogner:2007rx}
Bogner S~K, Furnstahl R~J, Maris P, Perry R~J, Schwenk A and Vary J~P
2008 {\em Nucl. Phys.\/} {\bf A801} 21

\bibitem{Sternberg:2008:ACI:1413370.1413386}
Sternberg P, Ng E~G, Yang C, Maris P, Vary J~P, Sosonkina M and Le H~V 2008
  {\em Proc. of the 2008 ACM/IEEE conference on Supercomputing\/} 
  (Piscataway, NJ, USA: IEEE Press) pp 15:1--15:12 ISBN 978-1-4244-2835-9

\bibitem{Vary:2009qp}
Vary J~P, Maris P, Ng E, Yang C and Sosonkina M 2009 {\em J. Phys. Conf. Ser.\/} {\bf 180} 012083

\bibitem{Maris201097}
Maris P, Sosonkina M, Vary J~P, Ng E and Yang C 2010 {\em Procedia Computer
  Science\/} {\bf 1} 97

\bibitem{Srivinasa2012256}
Srinivasa S, Sosonkina M, Maris P, and Vary J~P 2012 {\em Procedia Computer 
  Science\/} {\bf 9} 256

\bibitem{Maris:2011as}
Maris P Vary J~P, Navratil P, Ormand W~E, Nam H, and Dean D~J
2011 {\em Phys. Rev. Lett.\/} {\bf 106} 202502

\bibitem{rotbands}
Bohlen H~G {\em et~al.\/} 2008 {\em J. Phys.: Conf. Ser.\/} {\bf 111} 012021

\bibitem{textbook}
Rowe D~J 1970 {\em Nuclear Collective Motion} (London: Methuen, London)

\bibitem{caprio2012}
Caprio M~A, Maris P and Vary J~P 2012 {\em in preparation}

\bibitem{Tilley2004155}
Tilley D, Kelley J, Godwin J, Millener D, Purcell J, Sheu C and Weller H 2004
  {\em Nuclear Physics A\/} {\bf 745} 155

\bibitem{Audi19971}
Audi G, Bersillon O, Blachot J and Wapstra A 1997 {\em Nuclear Physics A\/} {\bf 624} 1

\bibitem{Forssen:2004dk}
Forssen C, Navratil P, Ormand W~E and Caurier E 2005 {\em Phys. Rev.\/} {\bf  C71} 044312

\bibitem{Aroua:2002xm}
Aroua S {\em et~al.\/} 2003 {\em Nucl. Phys.\/} {\bf A720} 71

\bibitem{Negret:2006zz}
Negret A {\em et~al.\/} 2006 {\em Phys. Rev. Lett.\/} {\bf 97} 062502

\bibitem{LSO}
Suzuki K and S. Y. Lee S Y 1980 {\em Prog. Theor. Phys.\/} {\bf 64} 2091;
Suzuki K 1982 {\em Prog. Theor. Phys.\/} {\bf 68}, 246; {\bf 68} 1999; 
Suzuki K and Okamoto R (1994) {\em Prog. Theor. Phys.\/} {\bf 92} 1045

\bibitem{Gazit:2008ma}
Gazit D, Quaglioni S and Navratil P 2009 {\em Phys. Rev. Lett.\/} {\bf 103} 102502

\bibitem{Vaintraub:2009mm}
Vaintraub S, Barnea N and Gazit D 2009 {\em Phys. Rev.\/} {\bf C79} 065501

\end{thebibliography}
%
\providecommand{\newblock}{}

\end{document}